\documentclass[12pt]{article}

\usepackage{multirow}
\usepackage{microtype}
\usepackage{graphicx}
\usepackage{subfigure}
\usepackage{booktabs} 
\usepackage{cleveref}
\usepackage{footmisc}
\usepackage{caption} 
\usepackage{hyperref}

\usepackage{amsmath}
\usepackage{amsfonts}
\usepackage{dsfont}
\usepackage{amssymb,amsthm,commath}
\usepackage{amsmath, amsthm, amssymb}
\usepackage{bm,xfrac,xcolor}
\usepackage{makecell}
\usepackage{nccmath}
\usepackage[toc,page,header]{appendix}

\usepackage{algorithm}
\usepackage[noend]{algorithmic}
\usepackage{bbm}
\usepackage{enumitem}
\usepackage{gensymb}
\usepackage{refcount}
\usepackage{mathtools}
\usepackage{stackengine}

\newcommand\samethanks[1][\value{footnote}]{\footnotemark[#1]}

\usepackage[top=1in,bottom=1in,left=1in,right=1in]{geometry}
\makeatletter
\def\myfnt{\ifx\protect\@typeset@protect\expandafter\footnote\else\expandafter\@gobble\fi}
\makeatother

\makeatletter
\def\BState{\State\hskip-\ALG@thistlm}
\makeatother





\hypersetup{colorlinks}

\def\1{\mathds{1}}

\newcommand{\Mcal}{{\mathcal{M}}}
\newcommand{\Ncal}{{\mathcal{N}}}

\newcommand{\Ucal}{{\mathcal{U}}}
\newcommand{\Vcal}{{\mathcal{V}}}

\newcommand{\Rmbb}{\mathbb{R}}

\newcommand{\embf}{\mathbf{e}}

\newcommand{\Rmbf}{\mathbf{R}}
\newcommand{\rmbf}{\mathbf{r}}

\newcommand{\umbf}{\mathbf{u}}

\newcommand{\vmbf}{\mathbf{v}}

\newcommand{\xmbf}{\mathbf{x}}

\newcommand{\ymbf}{\mathbf{y}}

\newcommand{\zmbf}{\mathbf{z}}

\newcommand{\rs}{\mathbf{r}^*}

\newcommand{\oline}[1]{\mkern 1.5mu\overline{\mkern-1.5mu#1}}

\renewcommand{\hbar}{\oline{h}}

\newcommand{\xdownarrow}[1]{%
  {\left\downarrow\vbox to #1\right.\kern-\nulldelimiterspace}
}

\DeclareMathOperator{\dom}      {dom}

\DeclareMathOperator{\argmin}   {arg\,min}

\newcommand{\baligned}     {\begin{aligned}}
	\newcommand{\ealigned}     {\end{aligned}}
\newcommand{\barray}       {\begin{array}}
	\newcommand{\earray}       {\end{array}}
\newcommand{\bbmatrix}     {\begin{bmatrix}}
	\newcommand{\ebmatrix}     {\end{bmatrix}}
\newcommand{\bcases}       {\begin{cases}}
	\newcommand{\ecases}       {\end{cases}}
\newcommand{\bcenter}      {\begin{center}}
	\newcommand{\ecenter}      {\end{center}}
\newcommand{\bcolumn}      {\begin{column}}
	\newcommand{\ecolumn}      {\end{column}}
\newcommand{\bcolumns}     {\begin{columns}}
	\newcommand{\ecolumns}     {\end{columns}}
\newcommand{\benumerate}   {\begin{enumerate}}
	\newcommand{\eenumerate}   {\end{enumerate}}
\newcommand{\bequation}    {\begin{equation}}
	\newcommand{\eequation}    {\end{equation}}
\newcommand{\bequationn}   {\begin{equation*}}
	\newcommand{\eequationn}   {\end{equation*}}
\newcommand{\bfigure}      {\begin{figure}}
	\newcommand{\efigure}      {\end{figure}}
\newcommand{\bflushright}  {\begin{flushright}}
	\newcommand{\eflushright}  {\end{flushright}}
\newcommand{\bitemize}     {\begin{itemize}}
	\newcommand{\eitemize}     {\end{itemize}}
\newcommand{\bpmatrix}     {\begin{pmatrix}}
	\newcommand{\epmatrix}     {\end{pmatrix}}
\newcommand{\bsubequations}{\begin{subequations}}
	\newcommand{\esubequations}{\end{subequations}}
\newcommand{\btable}       {\begin{table}}
	\newcommand{\etable}       {\end{table}}
\newcommand{\btabular}     {\begin{tabular}}
	\newcommand{\etabular}     {\end{tabular}}
\newcommand{\bvmatrix}     {\begin{vmatrix}}
	\newcommand{\evmatrix}     {\end{vmatrix}}

\newcommand{\bequali}      {\bsubequations\begin{align}}
	\newcommand{\eequali}      {\end{align}\esubequations}

\newtheorem{assumption}{Assumption}
\newtheorem{prop}{Proposition}
\newtheorem{example}{Example}
\newtheorem{remark}{Remark}
\newtheorem{theorem}{Theorem}
\newtheorem{corollary}{Corollary}
\newtheorem{definition}{Definition}
\newtheorem{lemma}{Lemma}

\newcommand{\balgorithm}  {\begin{algorithm}}
	\newcommand{\ealgorithm}  {\end{algorithm}}
\newcommand{\balgorithmic}{\begin{algorithmic}}
	\newcommand{\ealgorithmic}{\end{algorithmic}}
\newcommand{\bassumption} {\begin{assumption}}
	\newcommand{\eassumption} {\end{assumption}}
\newcommand{\bcorollary}  {\begin{corollary}}
	\newcommand{\ecorollary}  {\end{corollary}}
\newcommand{\bdefinition} {\begin{definition}}
	\newcommand{\edefinition} {\end{definition}}
\newcommand{\bexample}    {\begin{example}}
	\newcommand{\eexample}    {\end{example}}
\newcommand{\bprop}    {\begin{prop}}
	\newcommand{\eprop}    {\end{prop}}
\newcommand{\blemma}      {\begin{lemma}}
	\newcommand{\elemma}      {\end{lemma}}
\newcommand{\bproblem}    {\begin{problem}}
	\newcommand{\eproblem}    {\end{problem}}
\newcommand{\bproof}      {\begin{proof}}
	\newcommand{\eproof}      {\end{proof}}
\newcommand{\bremark}     {\begin{remark}}
	\newcommand{\eremark}     {\end{remark}}
\newcommand{\btheorem}    {\begin{theorem}}
	\newcommand{\etheorem}    {\end{theorem}}
\usepackage{placeins,afterpage}
\usepackage[customcolors]{hf-tikz}
\usepackage{adjustbox}
\hypersetup{colorlinks,urlcolor=blue, citecolor=black}

\renewcommand{\text}[1]{{\mathrm{#1}}}

\providecommand{\norm}[1]{{ \left\lVert{#1}\right\rVert }}

\newcommand{\vect}[1]  {{\boldsymbol{#1}}}

\def\va{{\vect{a}}}
\def\vb{{\vect{b}}}

\def\vw{{\vect{w}}}

\def\bb{{\mathbf{b}}}

\def\bb0{{\mathbb{0}}}
\def\bb1{{\mathbb{1}}}

\def\bbR{{\mathbb{R}}}

\DeclareMathOperator{\Int}{int}

\newcommand{\indom}[1] {\dom(#1) }
\newcommand{\inint}[1] {\Int(#1) }

\newcommand{\conj}[1]  {\left({#1}\right)^*}
\newcommand{\inv}[1]   {{\left(#1\right)}^{-1}}
\newcommand{\dotp}[2]  {\left\langle {#1}, {#2} \right\rangle }

\newcommand{\trans}[1] {{{#1}^\top}}

\newcommand{\kl}[2]  {\mathrm{KL} \left( {#1} \Vert  {#2} \right)  }

\newcommand{\gi}[2]  {\mathrm{GI} \left( {#1} \Vert  {#2} \right)  }

\newcommand{\dphi}[2]{D_{\vect{\phi}} \left( {#1} ,  {#2} \right)  }

\newcommand{\Rp} {{\mathbbm{R}_+}}
\renewcommand{\S}{\Delta}

\newcommand{\sij}  {w_{ij}}

\renewcommand{\u}  {\vect{u}}
\renewcommand{\v}  {\vect{v}}

\newcommand{\vr}   {\vect{r}}

\newcommand{\rij}  {R_{ij}}
\newcommand{\trij} {\hat{R}_{ij}}

\newcommand{\vs}   {{|\Vcal_i|}}

\newcommand{\mRL}   {{\mathrel\mathcal{R}_L}}
\newcommand{\mmRL}  {{\mathrel\mathcal{R}_{L, \epsilon}}}

\newcommand{\mRLp}   {{\mathrel\mathcal{R}^+_L}}
\newcommand{\mmRLp}  {{\mathrel\mathcal{R}^+_{L, \epsilon}}}

\renewcommand{\dphi}[2]{D_\phi \big( {#1} \bigl|\bigl| \bigr.  {#2} \big)  }

\newcommand{\dgphi}[2]  {\db{\phi}{#1}{\inv{\nabla \phi} \left( {#2}
\right)}}
\newcommand{\lphi}[2] {C_{\phi} \big( {#1} \bigl|\bigl| \Bigr.  {#2} \big)}

\newcommand{\db}[3]{D_{#1} \big( {#2} \bigl|\bigl| \bigr.  {#3} \big)}

\newcommand{\vx}  {\vect{x}}

\usetikzlibrary{calc}
\begin{document}

\title{Clustered Monotone Transforms for Rating Factorization}
\author{Gaurush Hiranandani\thanks{Equal Contribution} \,\thanks{University of Illinois Urbana-Champaign}\\
gaurush2@illinois.edu
\and
Raghav Somani\samethanks[1] \,\thanks{Microsoft Research} \\
t-rasom@microsoft.com
\and
Oluwasanmi Koyejo\samethanks[2]\\
sanmi@illinois.edu
\and
Sreangsu Acharyya\samethanks \\
srach@microsoft.com
}
\date{\today}

\flushbottom
\maketitle

\begin{abstract}
Exploiting low-rank structure of the user-item rating matrix has been the crux of many recommendation engines. However, existing recommendation engines force raters with heterogeneous behavior profiles to map their intrinsic rating scales to a common rating scale (e.g. 1-5). This non-linear transformation of the rating scale shatters the low-rank structure of the rating matrix, therefore resulting in a poor fit and consequentially, poor recommendations. In this paper, we propose Clustered Monotone Transforms for Rating Factorization (CMTRF), a novel approach to perform regression up to unknown monotonic transforms over unknown population segments. Essentially, for recommendation systems, the technique searches for monotonic transformations of the rating scales resulting in a better fit. This is combined with an underlying matrix factorization regression model that couples the user-wise ratings to exploit shared low dimensional structure. The rating scale transformations can be generated for each user, for a cluster of users, or for all the users at once, forming the basis of three simple and efficient algorithms proposed in this paper, all of which alternate between transformation of the rating scales and matrix factorization regression. Despite the non-convexity, CMTRF is theoretically shown to recover a unique solution under mild conditions. Experimental results on two synthetic and seven real-world datasets show that CMTRF outperforms other state-of-the-art baselines.
\end{abstract}
\section{Introduction}
\label{sec:intro}

Identifying and exploiting the low-rank structure in a rating matrix \cite{singh2008unified,koren2009matrix, yang2014survey, bokde2015matrix} has been the basis of many effective recommendation algorithms. It is believed that the low-rank structure originates from the existence of different {\emph{factors}} that users evaluate an item on. The numeric user-rating of an item is the weighted sum of the strengths with which the item possesses the factors, weighted by the user-specific factor importance. Naturally, this spans all real numbers. Such low-rank structure is, however, destroyed by nonlinear transformations. This paper is about increasing the predictive power of low-rank models by means of accounting for the disruptions from nonlinear transformations of the original user-ratings.

Recommendation engines that solicit user-ratings usually constrain the user-ratings to a fixed numerical scale. This scale is often a discrete subset of real numbers. For instance, GoodReads\footnote{www.goodreads.com} uses a rating scale of 1-5, whereas IMDB\footnote{www.imdb.com} has a rating scale of 1-10. However, users often have individualized scales which express their strength of `likes' and `dislikes'. In order to rate items, a user has to map and quantize her personal scale to the common scale offered by the recommendation engine, leading to the aforementioned wrecking of the low-rank structure due to these non-linear transformations. A good illustration is shown in Figure 1 of Ganti et al.~\cite{ganti2015matrix}, implying that traditional matrix completion / factorization methods may be ineffective even in the presence of mild nonlinearities.

The heterogeneity in user rating scales is an issue that has been noted by others \cite{beutel2014cobafi}. However, to our knowledge, no solution has been proposed to accommodate this from the perspective of rating scales. Our empirical results suggest that appropriately modeling user-heterogeneity improves performance because  
low-rank models are more effective when applied to the unknown, innate user-ratings, i.e., prior to mapping those to the engine's scale. We improve upon state of the art techniques based on this point of view. In particular, we address the (non-linear) scale mismatch between users and the engine that adversely affects the low-rank structure.

To preserve the user's relative preference over the rated items, the mapping from the user-space rating to the engine's rating has to be monotonic. With this observation in mind, we search the space of all (strongly) monotonic link functions (Section \ref{sec:background}) that transform the user's innate ratings into those that are recorded in the engine. In our paper, we seek a monotonic transformation that obtains the `best' low-rank structure, where we use the Mean Square Error (MSE) criterion to quantify what is `best'. Although our empirical results focus on MSE, we show that the techniques are applicable to Bregman Divergences as well, which are a much broader class of loss functions applied to measuring ranking quality \cite{ravikumar2011ndcg, acharyya2012learning}.

In this paper, we explore the cases: (i) where every user has her own monotonic transformation of the rating scale, (ii) where there is a common monotonic transformation of the rating scale that applies to all the users, and lastly and more realistically, (iii) where clusters of users share the same monotonic transformation of the rating scale. 
It is worth emphasizing that in case (iii) -- which is the main thrust of our paper -- effectively, it is the monotonic functions that get clustered. Both the monotonic transformation(s) and the clustering are a-priori unknown and are obtained from the data. Note that cases (i) and (ii) are the extreme ends of the spectrum: the former risks overfitting whereas the latter risks an underfit. 
Before diving into further technical details, we summarize the main contributions of this paper as follows:
\bitemize[leftmargin=0.2in]
  \item We propose Clustered Monotone Transforms for Rating Factorization (CMTRF), a user-personalized rating transformation model for collaborative filtering. CMTRF jointly searches over monotonic transformations of rating scales for each user (or cluster of users) and applies matrix factorization regression that exploits shared structure in the user-item rating matrix.
\item We propose three simple and efficient algorithms: \texttt{CMTRF}, \texttt{1-CMTRF}, and \texttt{N-CMTRF}. The first one takes the number of clusters $K$ as input and corresponds to $K$ different transformations for $K$ clusters of users. The latter two are special cases corresponding to one transformation for all users and $N$ different transformations for $N$ users, respectively. All the algorithms involve simple optimization scheme that alternates between the rating scale transformation step and the regression step.
\item We show that when squared loss is used as the regression loss function, CMTRF has a unique solution in terms of transformed rating scale and regression matrix under verifiable conditions.
\item CMTRF is evaluated on two synthetic datasets and seven real-world benchmark recommendation datasets, where it consistently outperforms the state-of-the-art baselines.
\eitemize
Rest of the paper is organized as follows. We review literature in Section~\ref{sec:related}. In Section~\ref{sec:background}, we discuss necessary background for Bregman Divergences and matrix factorization. Our model and the resulting properties are discussed in  Section~\ref{sec:CMTRF}. In Section~\ref{sec:algorithms}, we propose our optimization schemes and algorithms. Experimental results are discussed in Section~\ref{sec:experiments}. We discuss more nuances and future work of the approach in Section~\ref{sec:discussion}. We conclude in Section~\ref{sec:conclusions}. 
\section{Related Work}
\label{sec:related}

Recommender systems is a mature field where theoretical and practical innovations continue to command strong interest, speaking volumes of their utility. Recommender systems have been applied to movies \cite{carrer2012social, winoto2010role}, music \cite{lee2010collaborative, tan2011using}, books \cite{nunez2012implicit}, documents \cite{porcel2012hybrid}, e-learning \cite{bobadilla2009collaborative}, e-commerce \cite{castro2011highly}, applications in markets \cite{costa2012app} and web search \cite{mcnally2011case}. Most widely used recommender systems have \emph{collaborative filtering} (CF) \cite{herlocker2004evaluating, su2009survey} at their heart. CF exploits ratings provided by users with similar tastes to offer new recommendations.

One of the most prominent approaches to perform collaborative filtering is Matrix Factorization (MF) \cite{koren2009matrix}. A unified approach for MF with Bregman divergences is proposed by Singh and Gordon \cite{singh2008unified}, showing that several well-known methods such as weighted singular value decomposition~\cite{kyrchei2017weighted}, maximum margin matrix factorization~\cite{srebro2005maximum}, non-negative matrix factorization~\cite{lee2001algorithms}, probabilistic latent semantic indexing~\cite{hofmann2017probabilistic} and other related models can be posed as special cases of this framework. Recommender systems are a demonstration that such richer models have utility.

Recently, there has been a surge in approaches that make use of monotonic link functions for the purpose of recommendations. We make note of two recent approaches. Retargeted Matrix Factorization (R-MF)~\cite{koyejo2013retargeted} learns to rank by searching for a monotonic transformation of the rating vector that results in a better fit while preserving the ranked order of each user's ratings. RMF finds a monotonic transformation of the entire rating vector (ratings provided to all the items) for all the users. Not only does this approach faces issues in dealing with monotonic transformations of large sparse rating vectors, whose sizes are inconsistent among users, it is also inapplicable to nonparametric regression with population segments, which makes it highly prone to overfitting. Further, R-MF is designed to solve the ranking problem; whereas, the proposed CMTRF solves the problem of predicting the exact ratings by the users and not just the order. Another approach is Monotonic single index model for Matrix Completion (MMC) \cite{ganti2015matrix} -- a related method that alternates between low-rank matrix estimation and monotonic function estimation that estimates missing matrix elements. MMC finds one monotonic transformation of the entire rating matrix making it prone to underfitting. Furthermore, in MMC, even elements with the same ratings may be mapped to different real numbers making the interpretation difficult. One of our algorithms, 1-CMTRF (discussed later), also finds one monotonic transformation of the rating scale, which is equivalent to one monotonic transformation of the rating matrix, but elements with the same ratings  are mapped to the same real number making the interpretation meaningful. Further, a significant way by which our proposed method, K-CMTRF, departs from both of these is that it partitions the users into groups that share their monotonic transformation of the rating scale. More importantly, our results highlight how user-specific and user-shared monotonicities are a useful inductive bias for recommendation systems -- outperforming both the individual and global approach.

We provide a novel approach to perform regression up to unknown monotonic transforms over unknown population segments. While fitting parametric models is routine, we illustrate how to search jointly and efficiently over monotonic functions and population segments. Recommender systems serve as an easy to motivate application familiar to the audience. The paper also shows how the results can be extended to Bregman divergences and related generalized linear models, both of which have a rich history.
\section{Notations and Background}
\label{sec:background}

We denote the set of real numbers by $\mathbb{R}$. Vectors are denoted by bold lower case letters, and matrices are capitalized. $\xmbf^T$ denotes the transpose of vector $\xmbf$. $\norm{\xmbf}$ and $\dotp{\cdot}{\cdot}$ denote the $\ell_2$ norm and the inner-product, respectively. 
A vector $\xmbf$ is defined to be in descending order if $x_i \geq x_j$ when $i < j$. We denote the index set of size $L$ by $[L]$. The set of vectors of size $L$ in decreasing order is denoted by $\mRL$. Vector $\xmbf$ is isotonic with $\ymbf$ if $x_i \geq x_j$ implies $y_i \geq y_j$.
The positive orthant for $d$ dimensions is denoted by $\Rmbb^d_+$, and $\Rmbb^d_\epsilon$ denotes its subset where each component is bounded away from $0$ by $\epsilon$.

{\bf Problem Setup:} Let $N$ and $M$ denote the number of users and items, respectively. Let $W$ be the binary matrix whose entries $\sij$ denote whether user $i$ has 
rated item $j$ ($\sij =1$) or not ($\sij =0$). The true and predicted ratings are denoted by $\rij \in [L]$ and $\trij \in \Rmbb$, respectively.
We denote the base rating vector of size $L$ in decreasing order as $\rs$. For example, if $L = 5$, then $\rs = (5,4,3,2,1)^T$. For convenience, we will focus on the common binary encoding, where a rating is represented by a vector $\embf \in \{0, 1\}^L$, so $e_{L-l+1} = 1$ iff $l^{th}$ rating is provided to the item, and $0$ otherwise. For example, an item with rating 2 on 1-5 rating scale will have $\embf = (0,0,0,1,0)^T$. 
Let $\Ucal$ and $\Vcal$ denote the set of users and items, respectively. 
Let $\Vcal_i$ denote the set of items rated by user $i$, and $\vs$ be its cardinality. We define a matrix $E_i \in \Rmbb^{\vs \times L}$ for user $i$, such that every row is one hot binary rating encoder for items in $\Vcal_i$.  

Next, we discuss divergence functions considered in this paper. Let $\phi : \Theta \mapsto \bbR, \; \Theta = \dom \phi \subseteq \bbR^d $ be a strictly convex, closed function, differentiable on $\Int{\Theta}$. The corresponding Bregman divergence  $\dphi{\cdot}{\cdot}: \indom{\phi} \times \inint{\indom{\phi}} \mapsto \Rp$ is defined as  $\dphi{\vect{x}}{\vect{y}} \triangleq \phi(\vect{x}) - \phi(\vect{y}) - \dotp{ \vect{x} -\vect{y}} {\nabla \phi(\vect{y})}.$ From strict convexity of $\phi$, it follows that $\dphi{\vect{x}}{\vect{y}} \geq 0$ and $\dphi{\vect{x}}{\vect{y}} = 0$ iff.
$\vect{x}=\vect{y}.$ Bregman divergences are (strictly) convex in their first
argument, but not necessarily convex in their second.

In this paper, we consider functions $\phi(\cdot): \bbR^n \ni \vect{x} \mapsto \sum_i w_i\phi(x_i)$ that are (by overloading notation) weighted sums of {\emph{identical}} scalar convex function $\phi$ applied to each component. We refer to this class as {\emph{weighted, identically separable}} (WIS) function class.
Squared loss (SL), Kullback-Leibler divergence (KL), and generalized I-divergence (GID) are members of this family (Table \ref{tab:WIS_bregman}).

\begin{table}
\centering
\caption{Squared Loss (SL), KL Divergence (KLD), and Generalized I-Divergence (GID) are examples of weighted, identically separable Bregman Divergences (BD).}
\begin{tabular}{|c|c|c|}
\hline
\textbf{BD} &
$\phi(\vect{x})$              &   $\dphi{\vect{x}}{\vect{y}}$
\\ \hline \hline
SL & $\frac{1}{2}||\vect{x}||_W^2$  &   $\frac{1}{2}||\vect{x} - \vect{y}||_W^2$
\\
\hline
KLD & $\thead{\sum_i w_ix_i \log x_i, \;
            \vect{x} \in \S}$   &   $\thead{w\kl{x}{y} =\sum_i w_ix_i\log\left(\frac{x_i}{y_i}\right)}$\\
\hline
\makecell{GID} & $\thead{\sum_i w_i(x_i \log x_i - x_i), \\ \vect{x} \in \Rmbb_+^d}$   &   $\thead{w\gi{x}{y} \\ = \sum_i w_i [ x_i\log\left(\frac{x_i}{y_i}\right)  -x_i + y_i]}$ \\ \hline
\end{tabular}
\label{tab:WIS_bregman}
\end{table}

The \emph{Legendre conjugate} $\psi(\cdot)$ of the function $\phi(\cdot)$ is
defined as
$\conj{\phi}(\vect{x}) \triangleq \psi(\vect{x}) \triangleq
\sup_{\vect{\lambda}} ( \dotp{\vect{\lambda}}{\vect{x}} -
\phi(\vect{\lambda})).$
Furthermore, we take $\phi(\cdot)$ to be a convex function of Legendre type \cite{rockafellar1997convex}, for which, $\conj{ \conj{}}(\cdot)= \phi(\cdot)$ and $\inv{ \nabla \phi(\cdot)} = \nabla \psi(\cdot)$, is a one to
one mapping.
Lastly, a function $h(\vect{x})$ is $\alpha$-strongly monotonic if it satisfies 
$$\dotp{h(\vect{x}_1)-h(\vect{x}_2)}{\vect{x}_1-\vect{x}_2}\geq\frac{\alpha}{2}||\vect{x}_1-\vect{x}_2||_2^2.$$

\subsection{Matrix Factorization}
\label{ssec:mf_reco}

We combine the idea of monotonic  transformations of the rating scales with a popular regression technique known as matrix factorization (MF) \cite{koren2009matrix}. MF remains the predominant baseline method in the literature, and it is a struggle to outperform it in fair comparisons, thus is the ideal model to investigate the user-heterogeneity issues. As shown later, this factorization model not only provides empirical advantages but also has theoretical guarantees. 
In MF, the predictions take the form:
\begin{equation} \label{eq:dot}
\trij = \dotp{\u_i}{\v_j}, 
\end{equation}
where $\umbf_i, \vmbf_j \in \Rmbb^d$ are the factors for user $i$ 
and item $j$, respectively. Further, let $U \in \Rmbb^{|\Ucal| \times d}$ with $U(i,:) = \umbf_i$ represent the collected user factors, and let $V \in \Rmbb^{|\Vcal| \times d}$ with~$V(j, :) = \vmbf_j$ represent the collected item factors. Note that this factorization constrains the rank of the matrix $\hat{R}$ to be upper-bounded by $d.$ 
In MF, the factors are learned by minimizing the squared Euclidean distance
between the actual ratings and the predicted ratings as:
\begin{equation} \label{eq:sqr-loss} 
\min_{U, V} 
\sum_{i \in \Ucal, j\in \Vcal} \frac{1}{2} \sij\left(\rij - \trij\right)^2.
\end{equation}
Given that each user has rated some items in the system, the idea is to predict (compute the score $\trij$) how the users would rate the items that they have not yet rated, such that the system can make recommendations to the users.

\subsection{Retargeted Matrix Factorization}
\label{ssec:retar_mf_reco}

We make note of a distinct but interesting technique in the literature -- Retargeted Matrix Factorization (R-MF)  \cite{koyejo2013retargeted}. Instead of recommending items through accurate score predictions, R-MF focuses on obtaining the optimal order of items  for recommendation purposes. The authors \cite{koyejo2013retargeted} adapt the list-wise learning to rank (LETOR) algorithm called monotone retargeting \cite{acharyya2012learning}, to the recommendation task. 
Here, given any loss function $D: \Rmbb \times \Rmbb \rightarrow \Rmbb_+$, define a
pointwise LETOR based MF problem by:
\begin{equation*}
\underset{U, V }{\min} \sum_i \sum_j
\sij D\Big(\rij, f\left( \dotp{\u_i}{\v_j} \right)\Big)
\end{equation*}
where {$f: \Rmbb \mapsto \Rmbb$} is some
regression function with parameters $\u_i$ and $\v_j.$ If only ranking (ordering) is concerned, then the above objective function is unnecessarily stringent; 
therefore, the authors \cite{koyejo2013retargeted} optimize the following alternative objective:
\begin{equation*}\min_{ U, V, \{\Upsilon_i \in \Mcal\}} \sum_i \sum_j \sij D
\Big(\rij, \Upsilon_i \circ f\left(\dotp{\u_i}{\v_j} \right)\Big),
\end{equation*}
where $\Upsilon_i : \Rmbb \mapsto \Rmbb$ and $\Mcal$ is the class of all  monotonic increasing transformations. Without loss of generality, one may apply the monotonic transform to $\rij$ instead and optimize the following:
\begin{equation*}\min_{ U, V, \{\Upsilon_i \in \Mcal\}} \sum_i \sum_j \sij D
\Big(\Upsilon_i \left(\rij\right), f\left(\dotp{\u_i}{\v_j} \right)\Big).
\end{equation*}
CMTRF, which we discuss next, uses the same intuition as R-MF; however, it focuses on non-parametric regression to solve the problem of predicting the exact ratings by the users and not just the order. Further, for a user $i$, R-MF focuses on finding a monotonic transformation of the rating vector of size $| \mathcal{V}_i |$, which varies across users and can potentially be equal to the number of items; whereas, CMTRF focuses on the more fundamental transformation of the base rating-scale $\mathbf{r^*}$ of fixed size $L$, which is consistent across users and the proposed algorithms. This differentiation in the transformation not only provides advantages from the implementation perspective but also has a meaningful interpretation in the form of rating scale transformations. In addition, CMTRF explores partitions of the users into groups that share same monotonic transformation of the rating scale, thus can be applied to nonparametric regression with populations segments; whereas, R-MF is not equipped to handle this.
\section{Clustered Monotonic Transforms for Rating Factorization}
\label{sec:CMTRF}

We propose CMTRF, which learns (one or multiple) monotonic transformations of the system's rating scale, thus achieving more accurate low-rank structure and better predictions. Essentially, during the learning process, each level in the system's rating scale is mapped to a different real number satisfying the imposed monotonicity conditions. 
The rating vector for a user $i$ can be expressed as $E_i\rmbf$, where $\rmbf \in \mRL$. Since $E_i\rmbf$ are the targets given to the regressor to fit, it is more robust to enforce separation between the components, i.e., maintain $r_{k} \geq r_{k+1} + \epsilon$ for $k \in \{1,...,L-1\}$. We indicate the set of all such $\epsilon$ separated decreasing ordered vectors by $\mmRL.$ We next characterize the sets $\mRL$ and $\mmRL$. 

For two rating vectors $\rmbf_1, \rmbf_2 \in \mRL$, the convex composition $\vect{r} = \alpha \vect{r}_1 + (1-\alpha) \vect{r}_2$ preserves isotonicity; therefore, $\rmbf \in \mRL$. Furthermore, $\alpha \vect{r} \in \mRL$ for any $\alpha \in \Rp$ and $\vect{r} \in \mRL$. Hence, the set $\mRL$ is a convex cone. Similarly, for two rating vectors $\rmbf_1, \rmbf_2 \in \mmRL$, the convex composition $\vect{r} = \alpha \vect{r}_1 + (1-\alpha) \vect{r}_2$ lies in $\mmRL$. However, for $\rmbf \in \mmRL$, $\alpha \vect{r} \in \mmRL$ only if $\alpha \geq 1.$ Hence, $\mmRL$ is a convex set and equivalent to a cone cut at its vertex. This characterization makes the problem computationally tractable because each set can be described entirely by its extreme rays.

Notice that the above characterization of the set of base rating vectors $\mRL$ is useful for Bregman Divergences such as the squared loss. For Bregman Divergences such as the generalized I-Divergence (Table \ref{tab:WIS_bregman}), in addition to $\mRL$ and $\mmRL$, we will need the set of all component-wise positive vectors that are in descending order, i.e. $\mRL \cap \Rmbb^L_+$ that we represent by $\mRLp.$ Similar to the set $\mmRL$ we will use a well-separated version of $\mRLp$ denoted by $\mmRLp$ that consists of vectors whose components are not only strictly positive and sorted in decreasing order but also satisfy $r_k \geq r_{k+1} + \epsilon$ for $k \in \{1,...,L-1\}.$ Depending on the Bregman Divergence, one may use the appropriate counterparts of $\mRL$ (e.g. $\mRLp$). 
Next, we discuss three formulations of our main objective of finding a better fit using monotonic transformations of the rating scales. For clarity, the formulations are presented in terms of $\mRL$ and $\mmRL$.

\subsection{One Transformation for all the Users}
\label{ssec:1-sigma}
Here, we find one monotonic transformation of the recommendation engine's rating scale across all users. We define an objective as a loss function based on the predicted and actual ratings as follows: 
\begin{align}\label{eq:os}
\min_{U, V, \{\rmbf \in \mRL\}} & \sum_{i=1}^N D\left(E_i\rmbf , f\left(V_i\u_i\right)\right). 
\end{align}
The choice of the loss function $D$ is discussed in Section \ref{ssec:cost}. 
$f(V_i\umbf_i)$ denotes a regression functions of the factors, and $\rmbf$ denotes the one single transformation of the base rating vector $\rmbf^*$ across all users. 

\subsection{Separate Transformation for each User}
\label{ssec:N-sigma}
Next, we consider the possibility that every user has a different rating scale that is forced to the fixed scale of recommendation engine; therefore, we look for separate transformations of the base rating vector $\rmbf^*$ for every user. The objective is defined as follows:
\begin{align}\label{eq:ns}
\min_{U, V, \{\rmbf_i \in \mRL\}_{i=1}^N} & \sum_i D\left(E_i\rmbf_i , f\left(V_i\u_i\right)\right). 
\end{align}
The above objective function is almost similar to \eqref{eq:os}, except that $\rmbf_i$ stands for separate transformation for each user $i \in [N]$.

\subsection{Separate Transformation for each Cluster}
\label{ssec:K-sigma}
The above two formulations are at the two extremes, which might underfit (Section \ref{ssec:1-sigma}) or overfit (Section \ref{ssec:N-sigma}) the problem at hand. A more realistic assumption is that the clusters of users share the same monotonic transformations; therefore, the following formulation can be treated as a middle ground. Suppose we have $K$ clusters of users. Let $\zmbf_i$ denote the one hot encoding for cluster assignment for the user $i$. We seek $K$ transformations of the base rating vector $\rmbf^*$ corresponding to $K$ clusters of users via the following objective:
\begin{align}\label{eq:ks}
\min_{
\substack{
U, V, \{\zmbf_i\}_{i=1}^N,\\\{\rmbf_k \in \mRL\}_{k=1}^K}} & \sum_{k=1}^K\sum_{i=1}^m z_{ik} D\left(E_i\rmbf_k , f\left(V_i\u_i\right)\right),
\end{align}
where $z_{ik}$ is the $k$-th element of the assignment vector $\zmbf_i$. Notice that, effectively, it is the monotonic functions that get clustered and users are assigned to the monotonic functions which represent their rating behavior the best. Both monotonic transformations and clustering are a-priori unknown and are obtained from the data.

\subsection{Cost function}
\label{ssec:cost}
In order to make the problems \eqref{eq:os}-\eqref{eq:ks} tractable, we need to make a good choice of the distance like function $D\left(\cdot, \cdot\right)$ and the regression function $f(\cdot)$. We choose $D\left(\cdot, \cdot\right)$ to be a Bregman divergence $\dphi{\cdot} {\cdot}$ as defined
in Section \ref{sec:background}, and $f\left(V_i\u_i\right) = \inv{\nabla \phi} \left(V_i \u_i\right)$. 
This choice transforms \eqref{eq:os}-\eqref{eq:ns} into bi-convex optimization problems over a product of convex sets, as stated formally in the following lemma:

\blemma
\label{lem:convex}
The objective defined in \eqref{eq:os}-\eqref{eq:ks} are convex in $\hat{\Rmbf}_i = V_i\umbf_i$. In addition, \eqref{eq:os}-\eqref{eq:ks} are convex in $\rmbf$, $\rmbf_i \; \forall \; i \in [N]$, 
 and $\rmbf_k \; \forall \; k \in [K]$,
respectively.
\elemma
\bproof By definition, we have 
\begin{align}
\dgphi{E_i\vr}{V_i\umbf_i} &=  \psi\left(V_i\umbf_i\right) +
\phi\left(E_i\vect{r}\right) - \dotp{E_i\vect{r}}{V_i\umbf_i}
\label{eq:bddef}
\end{align}
where $\psi$ is the Legendre conjugate of $\phi$. Recall, the {Fenchel-Young} inequality~\cite{rockafellar2015convex}:
$\psi(\vw) + \phi(\vx) - \dotp{\vw}{\vx} \geq 0.$
Using Legendre duality, we see that~\eqref{eq:bddef} quantifies the gap in the Fenchel-Young inequality. As $\phi$ is WIS, clearly \allowbreak $\dgphi{E_i\vr}{V_i\umbf_i}$ is convex in $V_i\umbf_i$ and $\rmbf$. Hence, the objective in \eqref{eq:os} is convex in  $V_i\umbf_i$ and $\rmbf$. Similarly, \eqref{eq:ns} is convex in $V_i\umbf_i$ and $\rmbf_i$  for $i \in [N]$, and \eqref{eq:ks} is convex in $V_i\umbf_i$ and $\rmbf_k$  for $k \in [K]$.
\eproof

Notice that \eqref{eq:ks} is non-convex in general due to clustering; however, it is 
bi-convex if the clusters are fixed.

Lemma~\ref{lem:convex} ensures separate convexity in $\vr_i$ (or, $\rmbf$) and $V_i\umbf_i$ but not the joint convexity, which becomes crucial to obtain a global optima even for a coordinate-wise minimization because our constraint set is a product of convex sets. 
This is where the following theorem from~\cite{acharyya2012learning} (Theorem 2) comes handy. 
\btheorem
\label{joint}
The gap in the Fenchel-Young inequality:
$\psi(\vect{y}) + \phi(\vect{x}) - \dotp{\vect{x}}{\vect{w}}$ for any twice
differentiable strictly convex $\phi(\cdot)$ with a differentiable conjugate
$\conj{\phi}(\cdot) = \psi(\cdot)$ is jointly convex if and only if
$\phi(\vect{x}) = c ||\vect{x}||^2$ for all $c>0.$
\etheorem

For practical purposes, we maintain an explicit representation of the factor matrices $U$ and $V$. With this change, the optimization problem is no longer convex \emph{w.r.t.} these factors. However, Proposition 5 from~\cite{abernethy2009new} 
show conditions under which all local minima of the factorized form are global and correspond to the same matrix $\hat{R} = U \trans{V}$. Using a trace norm regularized version, maintaining the factorized form can be avoided. We trade running time and memory requirements for convexity of the cost function.

\begin{prop}
\label{localglobal}
(~\cite{abernethy2009new}: Proposition 5)
Let G be a twice differentiable convex function on matrices of size $p \times q$
with compact level sets. Let $d>1$ and $(U, V) \in \Rmbb^{p \times d} \times \Rmbb^{q
\times d}$ a local optimum of the function $H:\Rmbb^{p \times d} \times \Rmbb^{q
\times d} \mapsto \Rmbb$ defined by $H(U, V) = G\left(U\trans{V}\right)$, i.e., $(U, V)$ such
that $\nabla H(U,V) = 0$ and the Hessian of $H$ at $(U, V)$ is positive
semi-definite. If $U$ or $V$ is rank deficient, then $N=U\trans{V}$ is a global
optimum of $G$, that is $\nabla G(N) = 0$.
\end{prop}
Due to Theorem~\ref{joint} and Proposition~\ref{localglobal}, under mild conditions, 1-CMTRF \eqref{eq:os} and N-CMTRF \eqref{eq:ns} using squared loss recover a unique solution. K-CMTRF~\eqref{eq:ks} guarantees only local optima. Further, Proposition~\ref{localglobal} applied to other Bregman divergences like generalized I-Divergence can only
provide local optimality guarantees.
\section{Algorithms}
\label{sec:algorithms}

In this section, we discuss three variants of Clustered Monotone Transforms for Rating Factorization (CMTRF). To simplify the notation, we define $\lphi{\va}{\vb} = \dgphi{\va}{\vb}$. The resulting cost function corresponding to \eqref{eq:ns} is given by:
\begin{equation}\label{eq:mrtmp}
\min_{U, V, \{\rmbf_i \in \mRL\}_{i=1}^N }\sum_i 
\lphi{E_i\rmbf_i}{V_i\u_i}.
\end{equation} 
Similar cost functions can be obtained for \eqref{eq:os} and \eqref{eq:ks}. Since we are going to discuss algorithms, we would like to highlight one of the most important properties of Bregman Divergences from the implementation perspective.

Suppose for user $i$, $\hat{\Rmbf}_i = V_i\umbf_i \in \Rmbb^{|\Vcal_i|}$ represents the predicted rating vector obtained after one of the regression steps while optimizing \eqref{eq:ns}. 
Further, let $\Omega_{il}$ represent the number of items which got rating $l$ from the user $i$, i.e., $\Omega_{il} = \big\lvert\left\lbrace\left(i, j\right) \mid R_{ij} = l, j \in \Vcal_i\right\rbrace\big\rvert,$ and let ${\bar{\hat{R}}_i^l} = \frac{1}{\Omega_{il}}\sum_{j:R_{ij}=l} \hat{R}_{ij}$. Then the following lemma is valid for all \emph{WIS} Bregman Divergences.
\blemma
\begin{equation}\label{eq:bd_avg}
\underset{\rmbf_i \in \mRL}{\argmin}\  \lphi{E_i\rmbf_i}{\hat{\Rmbf}_i} = \underset{r_{i1} < \cdots < r_{iL}}{\argmin}\ \sum_{l=1}^L \Omega_{il}\lphi{r_{il}}{\bar{\hat{R}}_i^l} \nonumber
\end{equation}
\label{lem:simple}	
\vspace{-3mm}
\elemma
\bproof
\begin{align}
& \lphi{E_i\rmbf_i}{\hat{R}_i}  
= \psi\left(\hat{R}_i\right) + \phi\left(E_i\rmbf_i\right) - \dotp{E_i\rmbf_i}{\hat{R}_i} \nonumber \\
&= \sum_{l=1}^L\Omega_{il}\Big[ \phi\left(r_{il}\right) - r_{il}  \sum_{j:R_{ij = l}} \frac{1}{\Omega_{il}} \hat{R}_{ij} + \sum_{j:R_{ij = l}} \frac{1}{\Omega_{il}} \psi\left(\hat{R}_{ij}\right)  \Big] \nonumber \\ 
&= \sum_{l=1}^L \Omega_{il} \Big[ \phi(r_{il}) - r_{il}  \sum_{j:R_{ij = l}} \frac{1}{\Omega_{il}} \hat{R}_{ij} \nonumber \\  
&+ \sum_{j:R_{ij = l}} \frac{1}{\Omega_{il}} \psi(\hat{R}_{ij}) + \psi\left(\frac{1}{\Omega_{il}}\hat{R}_{ij}\right) - \psi\left(\frac{1}{\Omega_{il}}\hat{R}_{ij}\right)  \Big] \nonumber \\
&= \sum_{l=1}^L \Omega_{il} \lphi{r_{il}}{\bar{R}_i^l} + \frac{1}{\Omega_{il}} \psi\left(\hat{R}_{ij}\right) - \psi\left(\frac{1}{\Omega_{il}}\hat{R}_{ij}\right) \nonumber 
\end{align}
\begin{align}
&\therefore\underset{\rmbf_i \in \mRL}{\argmin}\  \lphi{E_i\rmbf_i}{\hat{\Rmbf}_i} \nonumber \\
&= \underset{r_{i1} < \cdots < r_{iL}}{\argmin} \Bigg[\sum_{l=1}^L \Omega_{il}\lphi{r_{il}} {\bar{\hat{R}}_i^l} + \frac{1}{\Omega_{il}} \psi\left(\hat{R}_{ij}\right) - \psi\left(\frac{1}{\Omega_{il}}\hat{R}_{ij}\right)\Bigg]  \nonumber \\
&= \underset{r_{i1} < \cdots < r_{iL}}{\argmin} \sum_{l=1}^L \Omega_{il}\lphi{r_{il}} {\bar{\hat{R}}_i^l} \nonumber \qedhere
\end{align}
\eproof
Lemma 2, in particular for squared loss, tells us that the search for $\rmbf_i \in \mRL$ reduces to the well-known problem of weighted-isotonic regression \cite{kyng2015fast}. 
The parameters to learn are the $L$-elements $r_{i1},...,r_{iL}$ of the vector $\rmbf_i$. The target corresponding to $r_{il}$ is the average of the predicted ratings (i.e. after a matrix factorization step) for the items which got actual rating $l$ from user $i$. The same holds while searching $\rmbf$ in \eqref{eq:os} except that for each $l \in [L]$, the target is taken to be the average over the entire matrix, i.e.,
${\bar{\hat{R}}^l} = \frac{1}{\Omega_{l}}\sum_{(i, j):R_{ij}=l} \hat{R}_{ij}$,
where, $\Omega_{l} = \big\lvert\left\lbrace\left(i, j\right) \mid R_{ij} = l\right\rbrace\big\rvert$. Similarly, while optimizing \eqref{eq:ks}, the targets are averaged corresponding to the particular cluster.

\subsection{\texttt{1-CMTRF}}
\label{ssec:1_sigma}

\begin{figure}[t] \centering  \fbox{ \parbox{0.9\columnwidth}{ {\small{
\begin{align}
&\vect{x}^{t+1} = \underset{\vect{x}\in \mmRL}{\argmin}\,
  \sum_i \lphi{ E_i \vect{x}}{V_i^t \u_i^t }  \label{eq:x1} \\
&U^{t+1}, V^{t+1} = \underset{ U, V }{\argmin}
 \sum_i \lphi{ E_i \vect{x}^{t+1}}{ V_i \umbf_i } \label{eq:mf1}
 \\
 & \hspace{15ex}+ \frac{\lambda_u}{2} \sum_i ||\u_i||^2 +
 \frac{\lambda_v}{2} \sum_j ||\v_j||^2
  \notag
\end{align}
       }} \vspace*{-5pt}
}}
\caption{\label{fig:os_algo} \texttt{1-CMTRF} Algorithm.}
\end{figure}

Note that the cost function \eqref{eq:os} is not invariant to scale. For example, squared loss, KL divergence, and generalized I-divergence are homogeneous functions of degree 2, 1, and 1 respectively. Therefore, the cost function \eqref{eq:os} can be curtailed just by scaling its arguments down, without actually learning the task. To remedy this, we constrain $\rmbf$ to lie in an appropriate closed convex set not only separated from the origin but also to a set of vectors whose adjacent components are separated from each other. Therefore, we search for $\rmbf$ in $\mmRL$, defined in Section \ref{sec:CMTRF}. Furthermore, in order to safeguard against overfitting, we add squared Frobenius norm regularization for the matrices $U$ and $V.$  After these modifications, we obtain the final formulation as: 
\begin{equation}\label{eq:os_final}
 \min_{\substack{U, V, \\ \rmbf \in \mmRL}}
 \sum_i \lphi{E_i\rmbf}{ V_i\u_i } + 
  \frac{\lambda_u}{2} \sum_i \norm{\u_i}^2 + \frac{\lambda_v}{2} \sum_j \norm{\v_j}^2
\end{equation}
where, $\lambda_u, \lambda_v$ are tunable parameters. 
The combined algorithm for \texttt{1-CMTRF} is given in \figref{fig:os_algo}. We cycle through the two update steps until convergence. The update \eqref{eq:x1} can be solved using many constrained convex optimization algorithms. We simplify \eqref{eq:os_final} using Lemma \ref{lem:simple}, which enable us to implement \eqref{eq:x1} using isotonic regression technique -- Pool-Adjacent-Violators Algorithm (PAVA) \cite{JSSv032i05}. Translation techniques in PAVA allow us to enforce margin between adjacent components, enabling the justified use of the set $\mmRL$. The user and item
factors in \eqref{eq:mf1} can be updated using any  matrix factorization algorithm \cite{singh2008unified}.

\subsection{\texttt{N-CMTRF}}
\label{ssec:N_sigma}

\begin{figure}[t] \centering \fbox{ \parbox{0.9\columnwidth}{ {\small{
\begin{align}
&\vect{x}_i^{t+1} = \underset{\vect{x}_i\in \mmRL}{\argmin}\,
  \sum_i \lphi{ E_i \vect{x}_i}{V_i^t \u_i^t } \;\; \forall i  \text{\; in \; parallel} \label{eq:xN} \\
&U^{t+1}, V^{t+1} = \underset{ U, V }{\argmin}
 \sum_i \lphi{ E_i \vect{x}_i^{t+1}}{V_i \umbf_i } \label{eq:mfN}
 \\
 & \hspace{15ex}+ \frac{\lambda_u}{2} \sum_i ||\u_i||^2 +
 \frac{\lambda_v}{2} \sum_j ||\v_j||^2
  \notag
\end{align}
       }} \vspace*{-5pt}
}}
\caption{\label{fig:ns_algo} \texttt{N-CMTRF} Algorithm.}
\end{figure}

Instead of learning one monotonic transformation of the rating scale, now we learn a separate monotonic transformation for each user. For this case, the cost function is extended from \eqref{eq:ns} as follows:
\begin{equation}\label{eq:ns_final}
\hspace{-4.5mm}\min_{\substack{U, V, \\ \left\lbrace\rmbf_i \in \mmRL\right\rbrace_{i=1}^N}}
\hspace{-1.8mm}\sum_i \lphi{E_i\rmbf_i}{ V_i\u_i }+\frac{\lambda_u}{2}\hspace{-0.5mm}\sum_i \norm{\u_i}^2+\frac{\lambda_v}{2}\hspace{-0.5mm}\sum_j \norm{\v_j}^2,
\end{equation}
where $\left\lbrace\rmbf_i\right\rbrace_{i=1}^N$ represent $N$ monotonic transformations -- one for each user. The algorithm to optimize \eqref{eq:ns_final} is shown in Figure \ref{fig:ns_algo}. Again, Lemma \ref{lem:simple} is used with targets being the averages corresponding to the user's ratings. Finding monotonic transformation for each user \eqref{eq:xN} can be embarrassingly parallelized and solved for each user separately using PAVA. Factor computations \eqref{eq:mfN} can be performed by any MF method \cite{singh2008unified}.   

\subsection{\texttt{CMTRF} Algorithm with K Clusters}
\label{ssec:k_sigma}

We denote the following algorithm by \texttt{CMTRF}, or simply by \texttt{K-CMTRF}, where $K \in \lbrace 2,...,N-1 \rbrace$ denote the number of clusters of users, which we tune in the algorithm. The idea is that the clusters of users can share same monotonic transformation of the base rating vector $\rmbf^*$. Restricting the search of monotonic transformation in $\mmRL$, the regularized cost function version of \eqref{eq:ks} is the following:
\begin{equation}\label{eq:ks_final}
 \min_{\substack{U, V, \left\lbrace\zmbf_i\right\rbrace_{i=1}^N \\ \left\lbrace\rmbf_k \in \mmRL\right\rbrace_{k=1}^K}}
 \sum_{k=1}^{K} \sum_{i=1}^{N} z_{ik} \lphi{E_i\rmbf_k}{ V_i\u_i } \\+ 
  \frac{\lambda_u}{2} \sum_i \norm{\u_i}^2 + \frac{\lambda_v}{2} \sum_j \norm{\v_j}^2
\end{equation}

The objective function in \eqref{eq:ks_final} suggests a natural iterative relocation algorithm given in Figure \ref{fig:ks_algo}. The algorithm takes number of clusters $K$ as input and assigns cluster $\zmbf_i^{t}$ \eqref{eq:clust} to each user $i$, as an intermediary step. Given $K$ monotonic transformations $\lbrace\xmbf_k^t\rbrace_{k=1}^K$ for the $K$ clusters and the predicted ratings from matrix factorization $V_i^t\umbf_i^t$, a user is assigned to a  cluster whose monotonic transformation gives the least Bregman Divergence. After cluster assignments, the optimization problem \eqref{eq:ks_final} is same as running $K$ instances of 1-CMTRF (Figure \ref{fig:os_algo}) for the $K$ clusters and solving \eqref{eq:x} and \eqref{eq:mf}, accordingly. For cluster initialization, we first run N-CMTRF (Figure \ref{fig:ns_algo}) and obtain $N$ transformations. 
Then, initial clusters are assigned and cluster centers are obtained using the K-means clustering algorithm \cite{lloyd1982least} over the $L$-length vector representations of the $N$ monotonic transformations. 
After the initialization, K-CMTRF (Figure \ref{fig:ks_algo}) is run to achieve the desired goal. Notice that K-CMTRF (Figure \ref{fig:ks_algo}) is similar to Bregman Hard Clustering algorithm~\cite{banerjee2005clustering} with a different re-estimation step~\eqref{eq:x}-~\eqref{eq:mf}. Hence, convergence to a local optimum is guaranteed.\footnote{See Proposition 2 and 3 of \cite{banerjee2005clustering} for the proof.}
\begin{figure}[t] \centering \fbox{ \parbox{0.9\columnwidth}{ {\small{
\begin{align}
&\zmbf_i^{t+1} = \underset{k \in [K]}{\argmin}\,
  \lphi{ E_i \xmbf_k^t}{V_i^t \u_i^t } \;\; \forall \; i \label{eq:clust} \\
&\vect{x}_k^{t+1} = \underset{\vect{x_k}\in \mmRL}{\argmin}\,
  \sum_{i: \zmbf_{ik}^{t+1}=1} \lphi{ E_i \vect{x_k}}{V_i^t \u_i^t } \;\; \forall k 
  \text{\; in \; parallel} \label{eq:x} \\
&U^{t+1}, V^{t+1} = \underset{ U, V }{\argmin}
 \sum_{k=1}^{K}\sum_{i: \zmbf_{ik}^{t+1}=1} \lphi{ E_i \vect{x_k}^{t+1}}{V_i \umbf_i } \label{eq:mf}
 \\
 & \hspace{15ex}+ \frac{\lambda_u}{2} \sum_i ||\u_i||^2 +
 \frac{\lambda_v}{2} \sum_j ||\v_j||^2
  \notag
\end{align}
       }} \vspace*{-5pt}
}}
\caption{\label{fig:ks_algo} \texttt{K-CMTRF} Algorithm with input $1<K<N$.}
\end{figure}
\section{Experiments}
\label{sec:experiments}
 
To show the efficacy of the proposed approach, we have conducted experiments with the following choices on both synthetic and real-world datasets. 

{\bf Experimental Choices:} We experimented with CMTRF using squared loss for two reasons: (a) the optimization theoretic guarantees discussed in Section~\ref{ssec:cost},  and (b) the baselines (discussed next) have been proven to be effective for metrics such as MSE. Further experiments with other Bregman Divergences will be covered in an extended version of this manuscript. We have compared our algorithm with four baselines: (a) regularized version of Matrix Factorization~\cite{singh2008unified}, (b) LMaFit~\cite{Wen2012}, which is a successive over-relaxation (SOR) based low rank factorization model, (c) Monotonic single index model for Matrix Completion (MMC)~\cite{ganti2015matrix}, which alternates between low-rank matrix estimation and monotonic function estimation to complete the matrix, and (d) Neural Network Matrix Factorization (NNMF)~\cite{dziugaite2015neural}, which replaces the inner product in the usual matrix factorization by a feed forward network that is learned from the data at the same time as the latent feature vectors. 
R-MF~\cite{koyejo2013retargeted} is designed to solve the ranking problem; hence, we do not compare our approach to it. For LMaFit, MMC, and NNMF, hyper-parameters were tuned according to the given references. For other algorithms, the regularization parameters $\lambda_u$ and $\lambda_v$ were tuned on the set $10^{\left\lbrace -2, -1.5, -1, \ldots, 2\right\rbrace}$. We chose $\epsilon = 0.5$. The parameter $K$ in $\texttt{K-CMTRF}$ was tuned on the set $\left\lbrace 2,3,5,10,20,30,50,75,100\right\rbrace$. Lastly, we used MSE and Mean Absolute Error (MAE) to compare our algorithms. For comparison with baselines, we evaluate the performance metrics by taking predictions as the inverse of the learned monotonic transformations. For instance, in 1-CMTRF, the final predictions $\hat{R}_{ij} = \Upsilon^{-1}\big(\dotp{\mathbf{u}_i}{\mathbf{v}_j}\big)$, where $\Upsilon$ is the inverse monotonic link function of $\rmbf$ \eqref{eq:os_final} obtained using linear spline interpolation.\footnote{\href{http://kluge.in-chemnitz.de/opensource/spline/}{http://kluge.in-chemnitz.de/opensource/spline}} 

\subsection{Synthetic Datasets and Results}
\label{ssec:synthetic}

\begin{table}[t]
	\centering
    \caption{MSE on synthetic datasets. 1-C, N-C, and K-C denote \texttt{1-CMTRF}, \texttt{N-CMTRF}, and \texttt{K-CMTRF}, respectively.}
	\begin{tabular}{|c|c|c|c|c|c|c|c|}
		\hline
		Data & MF & LMaFit & MMC & NNMF & 1-C & N-C & K-C \\ \hline
		SD-1 & 0.140 & 0.306 & 0.139 & 0.137 & 0.122 & \textbf{0.122} & 0.123\\
        SD-2 & 0.804 & 0.674 & 1.995 & 1.893 & 0.326 & 0.347 & \textbf{0.326} \\
		\hline
	\end{tabular}
	\label{tab:synth_mse}
\end{table}

\textbf{Datasets:} In synthetic datasets, we assumed each user has her own monotonic unrestricted rating scale isotonic to $\mRL$, where we chose $L=5$. We created two synthetic datasets: SD-1 and SD-2 as follows. First, we constructed two matrices $U$ and $V$ which contain i.i.d. samples from a Gaussian distribution (standard notation: $\Ncal\left(\mu, \sigma\right)$). Then, a low rank matrix  $Z= UV^T$ of size $3000 \times 2000$ is constructed from user and item factor matrices. In SD-1, for each user $i$, we consider a transformation $\Upsilon_i$ such that $\Upsilon_i(1) \sim \mathcal{N}\left(1,1\right)$ and $\Upsilon_i(i+1)= \Upsilon(i) + s,\; s\sim \Ucal\left(0.5,2\right)$ for $i \in \left\lbrace 2,3,4,5\right\rbrace$, where $\Ucal\left(a,b\right)$ is a uniform distribution on the interval $\lbrack a,b\rbrack$. This ensures the minimum gap constraint of $\epsilon = 0.5$ in the rating scale elements. The final rating matrix $X$ is constructed by transforming each row of the matrix $Z$, by applying $\Upsilon_i^{-1}$ element by element, i.e., $X(i, :) = \Upsilon_i^{-1}\left(Z(i, :)\right)$ for each row $i \in \lbrack 3000\rbrack$. In the other dataset, SD-2, each row is transformed using a monotonic function $\Upsilon_i(x) = -\frac{1}{c_i}\log\left(-1+\frac{5}{x-0.5}\right)$. The parameter $c_i$ is sampled from a Gaussian distribution. Notice that in both datasets, transformations are monotonic, however, arbitrary. 

\textbf{Results:} MSE values are shown in Table \ref{tab:synth_mse}. CMTRF beats the baselines on synthetic datasets. 
This means that if users do map their innate rating scales to recommendation engine's scale, then the proposed approach is effective. Next, we show results on real-world datasets, where most of our assumptions do not hold exactly.

\subsection{Real Dataset Experiments and Results}
\label{ssec:real}

 {\bf Datasets:} We used seven publicly available benchmark datasets: 
\begin{itemize}[leftmargin=*]

\begin{table}[t]
	\centering
    \caption{Datasets description.}
	\begin{tabular}{|c|c|c|c|c|c|}
		\hline
		Datasets & Users & Items & Ratings & Density & Split\\ \hline
		 ML100k & 751 & 1616 & 82,863 & 6.83\% & \multirow{5}{*}{\shortstack[l]{Chrono-\\logical}} \\
		ML1M & 5301 & 3682 & 901,851 & 4.62\% & \\
		ML10M & 62007 & 10586 & 6,950,602 & 1.06\% & \\
        GB & 42813 & 9403 & 4,729,637 & 1.17\% & \\
        Epinions & 77264 & 150497 & 808,690 & 0.007\% & \\ \hline
         Douban & 2999 & 3000 & 136891 & 1.52\% & \multirow{3}{*}{Uniform}\\
        Flixster & 2307 & 2945 & 26173 & 2.01\% & \\
        ML100k\_u & 943 & 1650 & 100000 & 6.43\% & \\
		\hline
	\end{tabular}
	\label{tab:datasets}
\end{table}

\item \textit{Movielens}\footnote{www.grouplens.org} is a movie recommendation website. We use movielens 100K (ML100k), movielens 1M (ML1M), and movielens 10M (ML10M) datasets.
First two datasets have ratings in $[5]$, and the third dataset has a rating scale of 0.5-5 with a step of 0.5.
\item \textit{Goodbooks}\footnote{www.fastml.com/goodbooks-10k \hspace{1.4cm}} is a book recommendation dataset. Ratings in Goodbooks (GB) take one value in the set $[5]$.
\item Epinions\footnote{\label{foot}www.trustlet.org/datasets/ \hspace{1.4cm}} is an item (such as products, companies, etc.) recommendation dataset.
Here, rating scale is 1-5, with a step of 1.
\item Douban\footnote{\label{fn:fmonti} https://github.com/fmonti/mgcnn} is a movie, book, music recommendation dataset. Here, ratings take one value in the set $[5]$.
\item Flixster\footref{fn:fmonti} is a movie based social website. Rating scale in Flixster is 0.5-5, with a step of 0.5.
\end{itemize}

{\bf Data preprocessing:} First, we remove all the users who have provided the same rating to all the rated items. Next, based on the train-test split (discussed next), we remove users who have ratings present in the test data but not in the train data. Same applies to the items as well. The number of users, items, and ratings after the pre-processing steps are provided in Table \ref{tab:datasets}. Notably, with an increase in number of users or decrease in number of ratings, the problem becomes challenging.

\textbf{Data Split:} We choose two strategies for train-test split: (a) Chronological Split: In ML100k, ML1M, ML10M, GB, and Epinions, first 80\% of the ratings based on timestamps are used for training, and the rest for testing. We choose chronological based split because rating data occurs in a timely fashion in real world; hence, a recommender system's performance should be evaluated accordingly, and (b) Uniform Split: To show a fair comparison to the reader, we also evaluate CMTRF on three common collaborative filtering benchmark datasets with standard uniformly chosen train-test split. These datasets are Douban, Flixster, and uniformly split Movielens 100K dataset (denoted by ML100K\_u).

\begin{table}[t]
	\centering
    \caption[Caption for LOF]{MSE and MAE on chronologically and uniformly split test datasets. 1-C, N-C, and K-C denote \texttt{1-CMTRF}, \texttt{N-CMTRF}, and \texttt{K-CMTRF}, respectively.}
	\begin{tabular}{|c|c|c|c|c|c|c|c|c|}
		\hline 
        \multicolumn{9}{|c|}{Chronological Split} \\ \hline \hline
		Metric & Data & MF & LMaFit & MMC & NNMF & 1-C & N-C & K-C \\ \hline 
		\multirow{5}{*}{MSE} & ML100k & 0.989 & 2.269 & 1.058 & 0.947 & \textbf{0.908} & 0.948 & 0.911 \\
		& ML1M & 0.809 & 1.918 & 0.989 & 0.797 & 0.799 & 0.790 & \textbf{0.778} \\
		& ML10M & 0.834 & 1.947 & 0.975 & 0.832 & 0.806 & 0.807 & \textbf{0.799} \\
        & GB & 0.811 & 3.232 & 0.888 & 0.768 & 0.778 & 0.772 & \textbf{0.757} \\
        & Epinions & 1.329 & 11.00 & - & \textbf{1.113} & 1.232 & 1.184 & 1.167 \\
		\hline
        \multirow{5}{*}{MAE} & ML100k & 0.989 & 1.173 & 0.830 & 0.755 & 0.759 & 0.758 & \textbf{0.752} \\
		& ML1M & 0.708 & 1.058 & 0.777 & 0.700 & 0.723 & \textbf{0.693} & 0.778 \\
		& ML10M & 0.712 & 0.988 & 0.791 & 0.698 & 0.708 & 0.697 & \textbf{0.691} \\
        & GB & 0.701 & 1.418 & 0.753 & 0.691 & 0.694 & 0.677 & \textbf{0.670} \\
        & Epinions & 0.864 & 2.886 & - & \textbf{0.800} & 0.845 & 0.812 & 0.811 \\
        \hline
        \multicolumn{9}{|c|}{Uniform Split} \\ \hline \hline
        Metric & Data & MF & LMaFit & MMC & NNMF & 1-C & N-C & K-C \\ \hline
		\multirow{3}{*}{MSE} & Douban & 0.543 & 1.462 & 0.846 & 0.542 & 0.541 & 0.534 & \textbf{0.534} \\
        & Flixster & 0.870 & 6.789 & 1.52 & 0.765 & 0.869 & 0.757 & \textbf{0.729} \\
        & ML100k\_u & 0.890 & 2.240 & 1.072 & 0.876 & 0.846 & 0.833 & \textbf{0.833} \\
		\hline
        \multirow{3}{*}{MAE} & Douban & 0.575 & 0.934 & 0.720 & 0.576 & 0.575 & 0.570 & \textbf{0.570} \\
        & Flixster & 0.704 & 2.005 & 1.015 & 0.667 & 0.710 & \textbf{0.623} & 0.630 \\
        & ML100k\_u & 0.749 & 1.160 & 0.808 & 0.744 & 0.728 & \textbf{0.720} & 0.721 \\
        \hline
	\end{tabular}
	\label{tab:mse_mae}
\end{table}

\textbf{Results:} Table \ref{tab:mse_mae} shows MSE and MAE values on test data for all the algorithms for both splits. First, we discuss results based on the chronologically split datasets. We observe that \texttt{1-CMTRF}, \texttt{N-CMTRF}, and \texttt{K-CMTRF} consistently outperform the baselines based on both the performance metrics. In particular, \texttt{K-CMTRF} performs the best for three out of five datasets.  
Unfortunately, the closest baseline to our work -- MMC, failed to run on larger and sparser dataset such as Epinions.\footnote{We ran out of $400$GB RAM on Epinions dataset because MMC uses a dense representation of the matrix.} 
Therefore, we could not report its performance on this dataset. Proposed algorithms work better than NNMF -- a non-linear approach for matrix factorization,  on four out of five datasets. 
NNMF works better than our approaches on only Epinions dataset, which is approximately a hundred times sparser dataset than ML10M. 
A possible reason attributed to this can be the sparsity of the dataset, which propels challenge in learning appropriate monotonic transformations. NNMF works better than MF, which in turn performs better than LMaFit. 

Moving on to the uniform split datasets, Table \ref{tab:mse_mae} shows that K-CMTRF achieves much lower MSE on Douban, Flixster, and ML100k\_u datasets compared to other baselines. The results are even better than many other baselines which make use of side information such as user and / or item graphs. Please see Monti et al.~\cite{monti2017geometric} for more baseline results.\footnote{Other baselines are reported in square root of MSE (RMSE).} We, on the other hand, just use the user-item rating matrix.  
These results show that the proposed approach not only has theoretically sound guarantees, but also performs better than state-of-the-art baselines with just a simple yet powerful and meaningful amalgamation of a novel idea.  

\section{Discussion and Future Work}
\label{sec:discussion}

\begin{figure*}[t]
\centering
\subfigure[ML100k]{\label{fig:mlsmall}\includegraphics[scale=0.35]{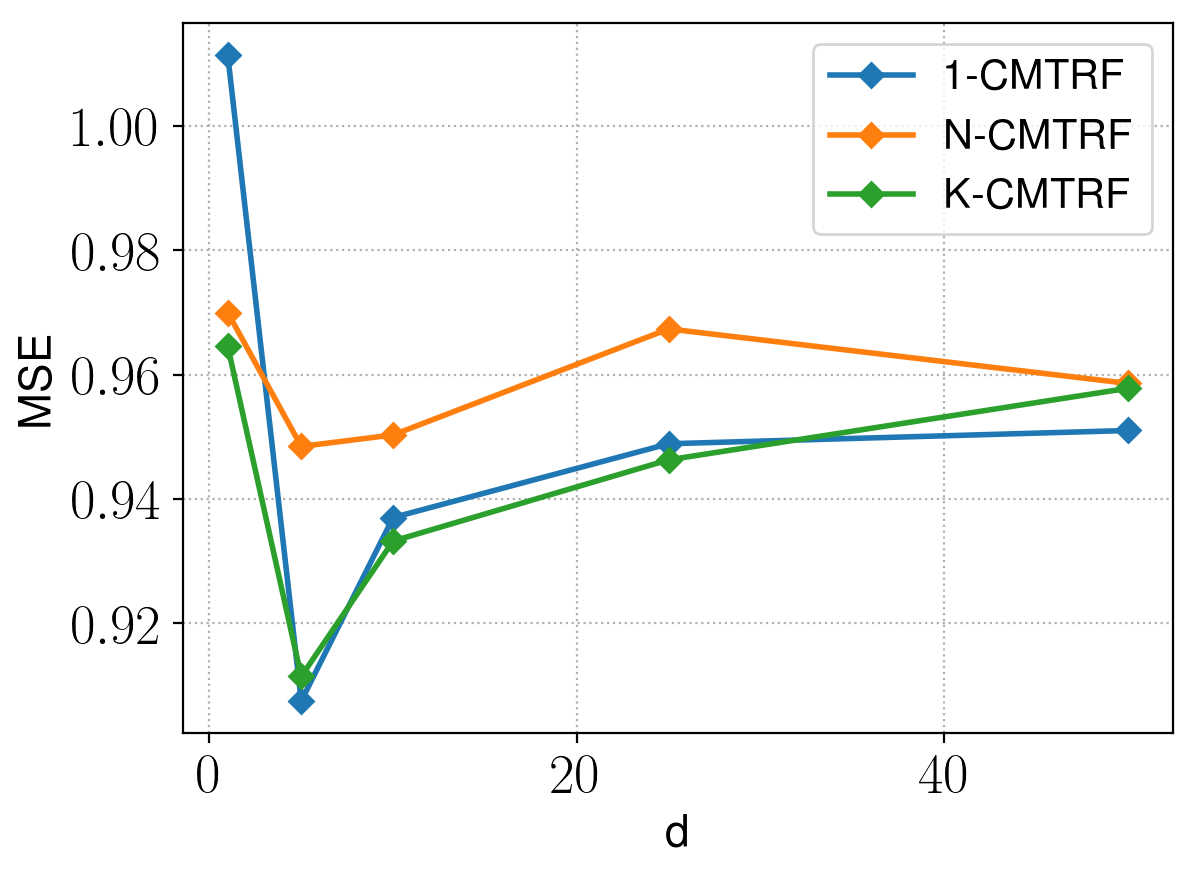}}
\subfigure[ML1M]{\label{fig:mlmedium}\includegraphics[scale=0.35]{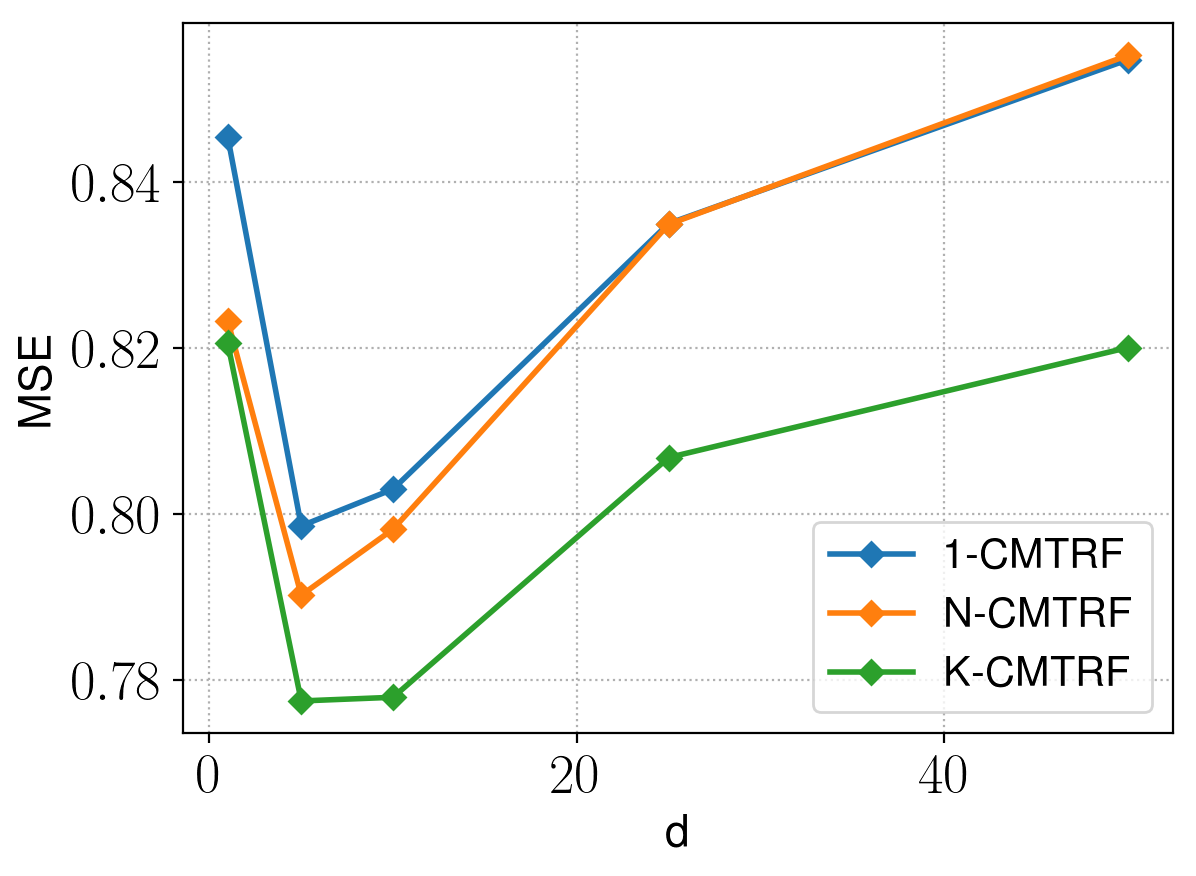}}
\subfigure[ML10M]{\label{fig:mllarge}\includegraphics[scale=0.35]{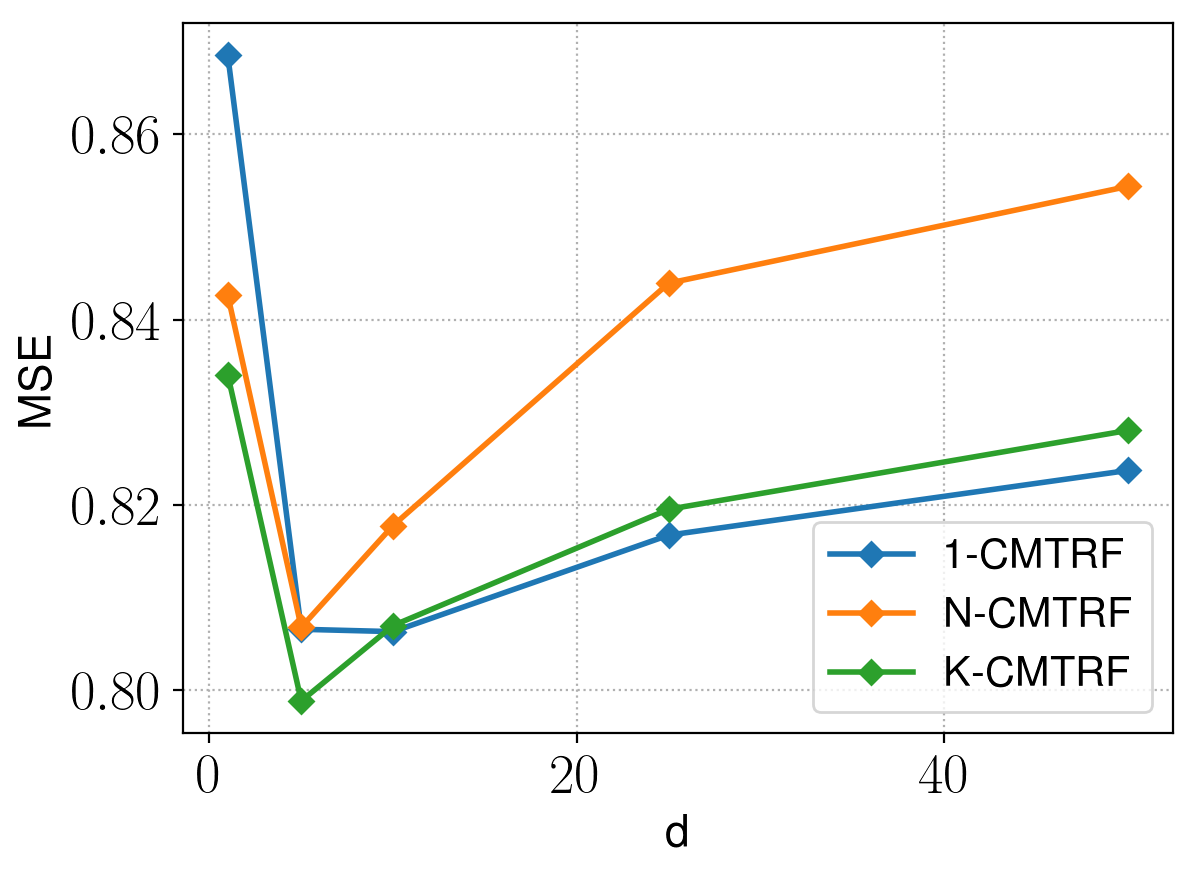}}
\subfigure[GoodBooks]{\label{fig:goodbooks}\includegraphics[scale=0.35]{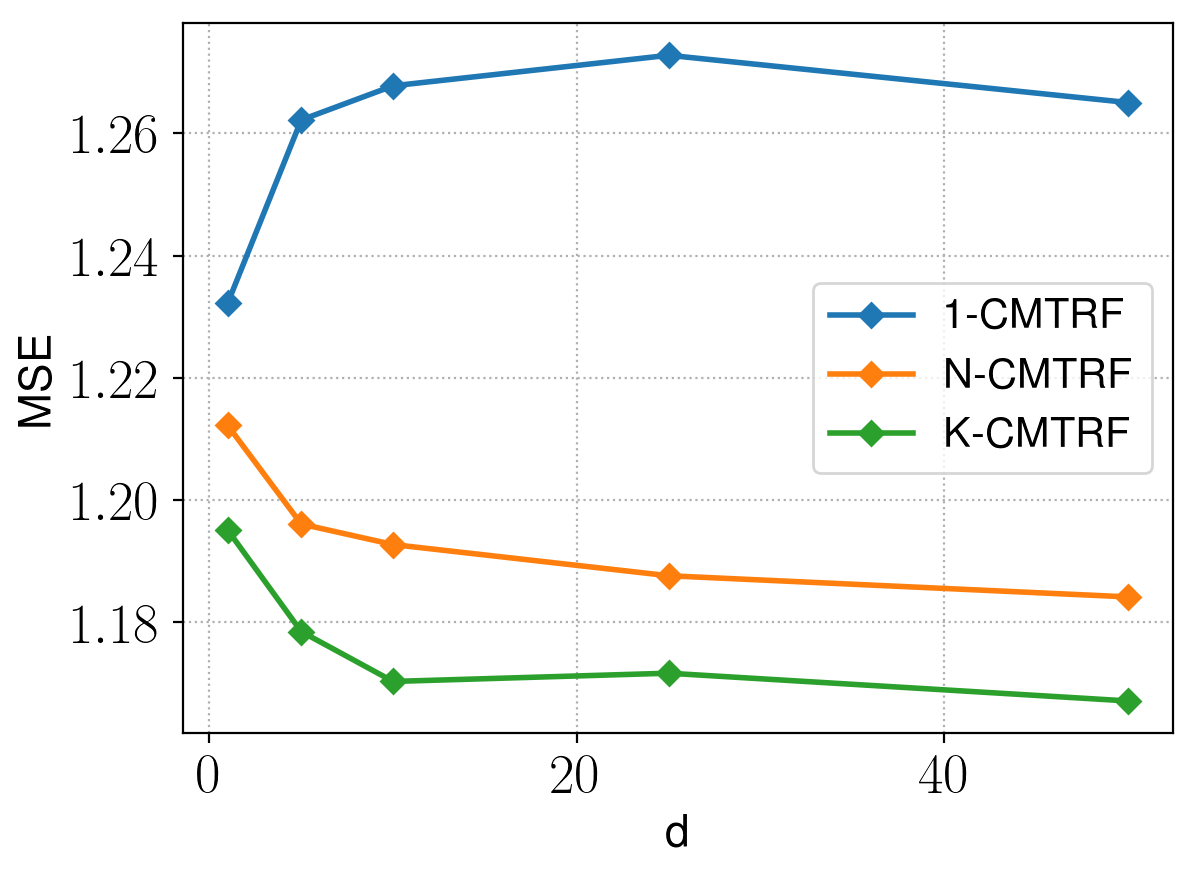}}
\subfigure[Epinions]{\label{fig:epinions}\includegraphics[scale=0.35]{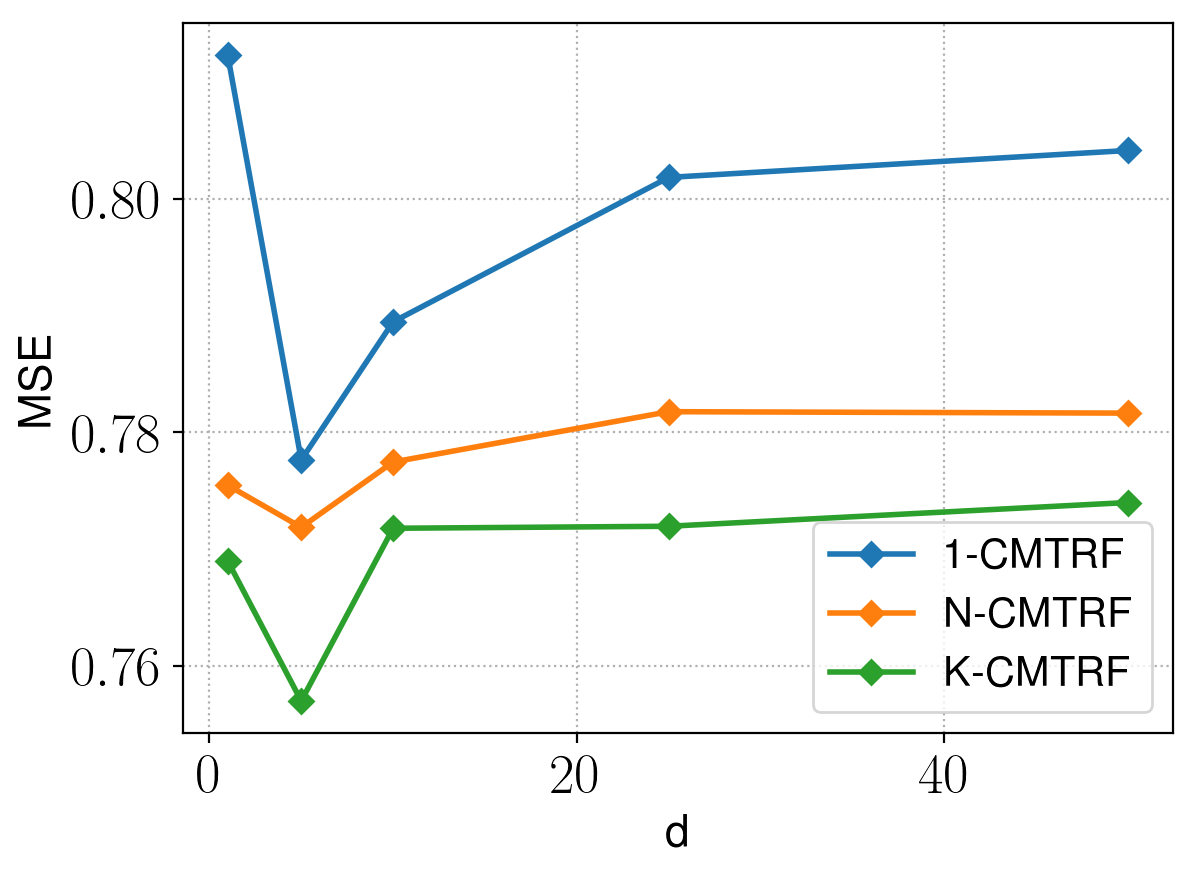}}
\caption{MSE of \texttt{1-CMTRF}, \texttt{N-CMTRF}, and \texttt{K-CMTRF} over factors (rank) $d$. Low-rank structure is observed in almost all the datasets.}
\label{fig:crossval}
\end{figure*}

\begin{figure}[t]
\centering
\subfigure[N-CMTRF Rating Scales]{\label{fig:mfrat}\includegraphics[scale=0.4]{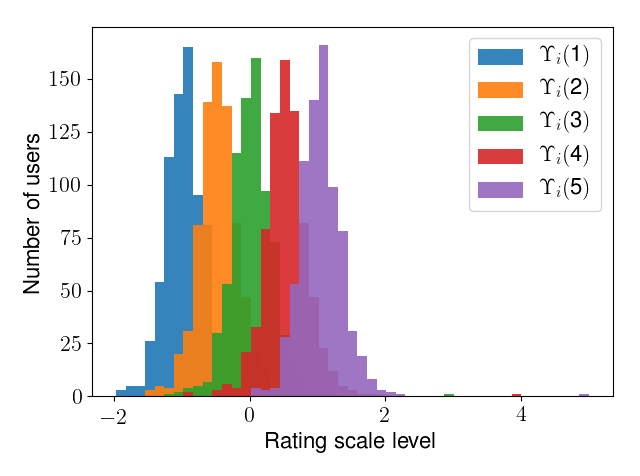}}
\subfigure[NNMF-N-CMTRF Rating Scales]{\label{fig:nnmfrat}\includegraphics[scale=0.4]{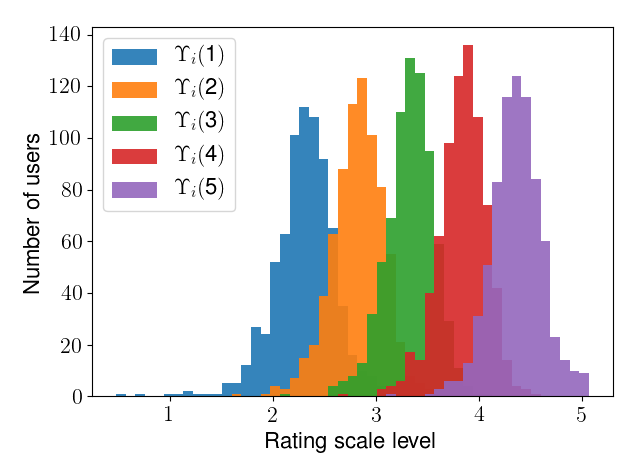}}
\caption{Transformed rating scales obtained for N-CMTRF and NNMF-N-CMTRF. Different histograms show transformation ($\Upsilon$) of different rating scale levels for all the users.}
\label{fig:ratings}
\end{figure}

Now, we discuss more nuances of CMTRF enabling us to understand its applications better. 
First, we highlight why clustered monotone transforms coupled with MF is good for CF. In Figure \ref{fig:crossval}, we show MSE across different number of factors $d$ (Section \ref{ssec:mf_reco}) for the proposed algorithms. Other parameters were tuned 
for each value of $d$. 
We observe that, in four out of five datasets, there is an optimal low-rank matrix 
where we achieve good MSE values. 
Next, we observed that the best $K$ (number of clusters) value for ML100k, ML1M, ML10M, Goodbooks, Epinions, Douban, Flixster, and ML100k\_u were 4, 3, 2, 100, 50, 20, 20, and 75, respectively. These numbers depend on our cross-validation set for $K$; however, practitioners can fix $K$ based on a-priori knowledge of user segments.

Lastly, we investigate the transformations of the rating scale on ML100k dataset. Figure \ref{fig:mfrat} shows monotonic rating scale transformations obtained for N-CMTRF algorithm. Different histograms are associated with different rating scale levels for all the users. We see that the latent scales lie in the range of $[-2,2]$, and the spread and peak of each histogram is almost equal for each rating scale level. Interestingly, in N-CMTRF, most of the learned rating scales lay on the boundary of the set $\mmRL$, i.e., for most of the users $r_k - r_{k+1} = \epsilon = 0.5$ for $k\in [4]$. This is because consecutive applications of monotonic transformation step \eqref{eq:xN} and U, V step \eqref{eq:mfN} forces the solution to lie on the boundary. Therefore, the mean of histograms are separated by approximately $\epsilon=0.5$. We observed the same trend, even when we put non-linear condition of the type $r_k - r_{k+1} = 1/\sqrt{k}$ for $k\in [4]$ due to the same reason. 
This suggests that factors adjust themselves with the provided monotonicity conditions. Nevertheless, different forms of enforcing monotonicity can be explored, which we leave for the future. 

Our proposal focuses on user-heterogeneity in the rating scales, thus 
in principle, can be combined with any recommendation system to enable flexible user-specific rating scales. We implemented a combination of NNMF and N-CMTRF, where the U, V steps \eqref{eq:mfN} in N-CMTRF algorithm (Figure \ref{fig:ns_algo}) are replaced by solution of NNMF.  
We call this implementation NNMF-N-CMTRF. 
Unfortunately, we faced many overfitting issues due to non-linear nature of the two steps. However, with the same monotonic constraints, we learned more non-linear rating scales. Figure \ref{fig:nnmfrat} shows rating scale transformation for the combined algorithm. The latent scales lie approximately in the range of $[2,5]$, and the spread and peak of each histogram is different for each rating scale level. Unlike N-CMTRF, this time most of the learned rating scales lay in the interior of the set $\mmRL$, i.e., $r_k - r_{k+1} > \epsilon = 0.5$ for $k\in[4]$. 
In future, we would combine rating scale transformation idea with other (non-linear) recommender systems with proper regularization and explore distances beyond Bregman Divergences. 
\section{Conclusion}
\label{sec:conclusions}
 
We propose Clustered Monotone Transforms for Rating Factorization (CMTRF), a novel 
approach to perform regression up to unknown monotonic transforms over unknown population segments. CMTRF improves rating predictions by searching over monotonic transformations of the rating scales for each user or a group of users. We propose three efficient algorithms - \texttt{CMTRF}, \texttt{1-CMTRF}, and \texttt{N-CMTRF}, suitable for transformations at various granularities of clusters of users. In addition, squared loss is shown to have nice optimization properties. We further show how the results can be extended to Bregman divergences and related generalized linear models. 
CMTRF is evaluated on synthetic and real-world datasets, where it consistently outperforms the baselines.
\bibliographystyle{plain}
\bibliography{references}
	
\end{document}